\documentclass[aps,pre,amsmath,amssymb,twocolumn,showpacs]{revtex4}     
\usepackage{epsfig}
\usepackage{graphicx}
\begin{document}

\title{Structural and dynamical heterogeneity in a glass forming liquid}

\author{Gurpreet S. Matharoo}
\affiliation{Department of Physics, St. Francis Xavier University,
  Antigonish, Nova Scotia B2G 2W5, Canada}
\author{M.S. Gulam Razul}
\affiliation{Department of Physics, St. Francis Xavier University,
  Antigonish, Nova Scotia B2G 2W5, Canada}
\author{Peter H. Poole} 
\affiliation{Department of Physics, St. Francis Xavier University,
  Antigonish, Nova Scotia B2G 2W5, Canada}

\date{\today}

\begin{abstract}
  We use the ``isoconfigurational ensemble'' [Phys. Rev. Lett. {\bf
    93}, 135701 (2004)] to analyze both dynamical and structural
  properties in simulations of a glass forming molecular liquid.  We
  show that spatially correlated clusters of low potential energy
  molecules are observable on the time scale of structural relaxation,
  despite the absence of spatial correlations of potential energy in
  the instantaneous structure of the system.  We find that these
  structural heterogeneities correlate with dynamical heterogeneities
  in the form of clusters of low molecular mobility.
\end{abstract}


\maketitle

Over the last decade, the identification and study of dynamic
heterogeneity (DH), especially in computer simulations, has added an
important new dimension to our understanding of complex relaxation in
glass forming liquids~\cite{E00,G00}.  DH refers to the emergence and
growth of spatially correlated domains of mobile and immobile
molecules as temperature $T$ approaches the glass transition
temperature $T_g$.  A question that dominates research on DH concerns
its connection to the structure of the liquid: What local
configurational properties influence whether a given molecule is
mobile or immobile?

Recent work by Widmer-Cooper, Harrowell and Fynewever~\cite{CHF04} has
shown conclusively that a structure-dynamics connection must exist at
the molecular level.  To do so, they use an ``isoconfigurational (IC)
ensemble''~\cite{CHF04,CH05,DH03}, a set of microcanonical molecular
dynamics (MD) trajectories in which each run starts from the same
initial equilibrium configuration, but with molecular momenta chosen
randomly accordingly to a Maxwell-Boltzmann distribution.  The result
is a set of trajectories lying on the same energy surface, and
evolving away from their common initial point in configuration space.
They then define and evaluate the ``dynamic propensity'': the average,
in the IC ensemble, of the squared displacement of a molecule at a
time comparable to the structural relaxation time.  They show that DH
is observed in this approach, in the form of increasing spatial
correlations of the dynamic propensities in a glass forming liquid as
$T \to T_g$.  Since the influence of the initial momenta is averaged
over, the observed spatial correlations must be configurational in
origin.

The strength of Ref.~\cite{CHF04} is that it exposed the features of
DH that are structural in origin, without needing to determine what
structural properties are responsible.  Other studies have worked
towards explicitly identifying structural correlators to dynamics.  A
number of works over the past decades have identified relationships
between average structural properties (especially free volume and
measures of symmetry in the local molecular environment) and bulk
dynamics; see e.g. Refs.~\cite{SNR81,H82,JA88,Starr02}.  More
recently, several studies have sought a correlation at the microscopic
level, e.g. between local free volume and local mobility, with more
success in some systems~\cite{LT06,JP05} than in others~\cite{CSW05}.
A notable absence of correlation between the local volume and the
local Debye-Waller factor has been reported recently~\cite{CHcm}.
Insights have also been realized using local measures of symmetry to
elucidate local mobility, e.g. icosohedral order~\cite{DSZ02,JP05}; or
the recent work of Shintani and Tanaka in which less mobile regions
were shown to correlate well to structured, crystal-like
domains~\cite{ST06}.  Recently, the local Debye-Waller factor has been
shown to correlate to the dynamic propensity~\cite{CH06}, firmly
establishing the connection between local dynamics at short and long
times.

Notably absent from the list of local structural quantities that
correlate well to local dynamics is the potential energy.  It has been
shown that a molecule with a low potential energy will be less mobile,
{\it on average}, than one with a high potential
energy~\cite{D99}. However, the variance around this average trend is
comparable to or larger than the trend itself, making it impossible to
predict what a {\it given} molecule will do based on its potential
energy.  Careful time averaging~\cite{LT06} and the use of inherent
structures~\cite{CH06} has been shown to yield little gain in
correlation.  This is both perplexing and disappointing.  Perplexing
since we know that the average potential energy strongly influences
the average dynamics of the system~\cite{NS04}.  And disappointing
because the potential energy is a natural observable to correlate to
dynamics, especially given the interest in the analysis of
glass-forming systems using the potential energy
landscape~\cite{Wales}.

In this Letter, by expanding the application of the IC ensemble to
structural quantities, we show that it is possible to identify
heterogeneities of the potential energy that correlate well to dynamic
heterogeneities, in a liquid where no useful correlation is
discernible from the instantaneous properties of the system.  Our work
shows that emergent ``structural heterogeneities'', that develop in
tandem with dynamic heterogeneities, exist and can be observed in a
glass forming liquid.

\begin{figure}
\centerline {\includegraphics[angle=270,width=3.0in]{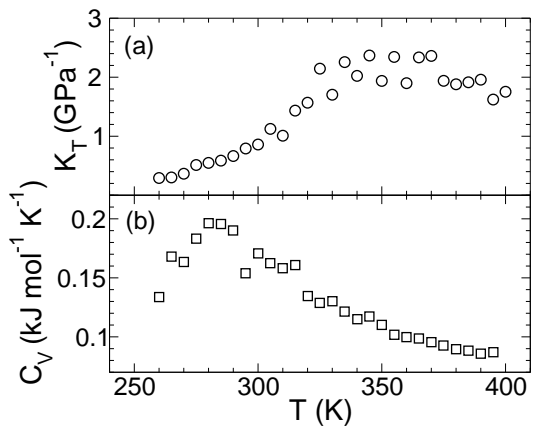}}
\caption{(a) $K_T$ and (b) $C_V$ as a function of $T$ along the
  $\rho=0.83$~g/cm$^3$ isochore.  Data are derived from the
  simulations described in Ref.~\protect{\cite{PSS05}}.}
\label{thermo}
\end{figure}

Our results are based on molecular dynamics simulations of $N=1728$
water molecules interacting via the ST2 pair
potential~\cite{st2}. Much is known about this simulation model, both
in terms of thermodynamic and transport behavior, making it a good
candidate for a detailed analysis of structure-dynamics relationships
in a model 3D molecular liquid~\cite{SPES97,PG99,PSS05,BPScm}.  We
study three $T$ (350, 290 and 270~K) all at density
$\rho=0.83$~g/cm$^3$.  At this $\rho$, the hydrogen bond network in
this water model is more prominent than at other $\rho$, suggesting
that local energetics may have a particularly strong influence on
dynamics.  This $\rho$ is also a convenient choice because the
isothermal compressibility $K_T$ (which is related to the strength of
static density fluctuations) is decreasing with $T$
(Fig.~\ref{thermo}). This ensures that any DH that emerges as $T$
decreases will not be due to the growth of conventional density
fluctuations.  At the same time, we note that the isochoric specific
heat $C_V$ (and hence the magnitude of fluctuations of the system
energy) is increasing to a maximum, so spatial variations of potential
energy may be occurring.  These conditions therefore provide a
promising regime in which to seek connections between energy and
dynamics at the molecular level.

In all our simulations, molecular interactions are cut off at a
distance of $0.78$~nm, and effects of those at longer range are
approximated via the reaction field method. All simulations have
constant $N$ and volume $V$, and use a 1~fs time step.  At each $T$,
we first conduct a standard MD run where $T$ is maintained near the
desired value using the method of Berendsen, et al~\cite{beren}.  To
ensure equilibration, these runs are carried out for twice the time
required for the mean squared displacement to reach 1~nm$^2$.  We then
use the configuration at the end of the equilibration run as the
initial configuration for generating $M=1000$ runs of an IC
ensemble. Note that both the linear and angular momentum of each
molecule is randomized, using the Maxwell-Boltzmann distribution
appropriate to the $T$ of the initial configuration.  Each run is
carried out in the microcanonical ensemble for a time $t=308$~ps (350
and 290~K) or $937$~ps (270~K).

\begin{figure}
\centerline {\includegraphics[width=3.9in]{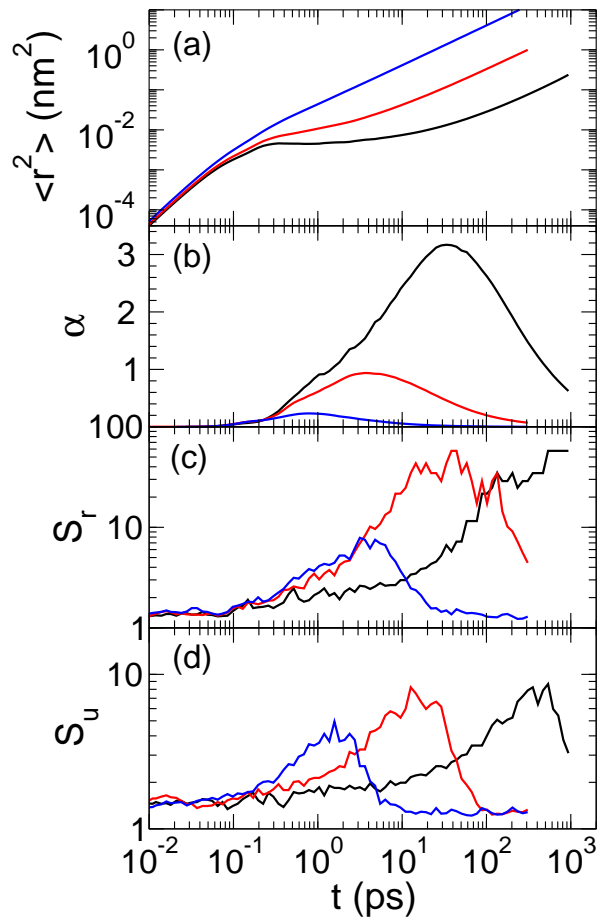}}
\caption{(a) Mean squared displacement $\langle r^2 \rangle$, (b)
  non-Gaussian parameter $\alpha$, (c) mean cluster size for least
  mobile molecules, $S_r$, and (d) mean cluster size for tightly bound
  molecules, $S_u$, all as a function of $t$.  From left to right in
  (a), the curves are for $T=350$, 290 and 270~K. In (b-d), the curve
  with the left-most maximum is $T=350$~K, the middle maximum is
  $290$~K, and the right-most is $270$~K.}
\label{ngp}
\end{figure}

Let $r^2(i,k,t)$ be the squared displacement of the O atom of molecule
$i$ at time $t$ in run $k$ of an IC ensemble.  The system-averaged and
IC-ensemble-averaged mean squared displacement, $\langle r^2 \rangle =
(NM)^{-1}\sum_{i=1}^N\sum_{k=1}^M r^2(i,k,t)$, and non-Gaussian
parameter, $\alpha=[(3\langle r^4\rangle)/(5 \langle r^2\rangle^2)]-1$
are shown in Fig.~\ref{ngp}.  Both show the characteristic pattern of
a glass forming liquid in which DH occurs.  $\langle r^2 \rangle$
develops a plateau at low $T$ indicating the onset of molecular
caging, and $\alpha$ displays an increasingly prominent maximum as $T$
decreases. We have checked that the curves in Figs.~\ref{ngp}(a-b)
are, at large $t$, within error of those found in a conventional
approach, where an average is taken over multiple time origins of a
single, long MD run.  This suggests that the particular initial
configurations chosen are at least approximately representative of the
equilibrium behavior.

\begin{figure}
\centerline {\includegraphics[width=3.2in]{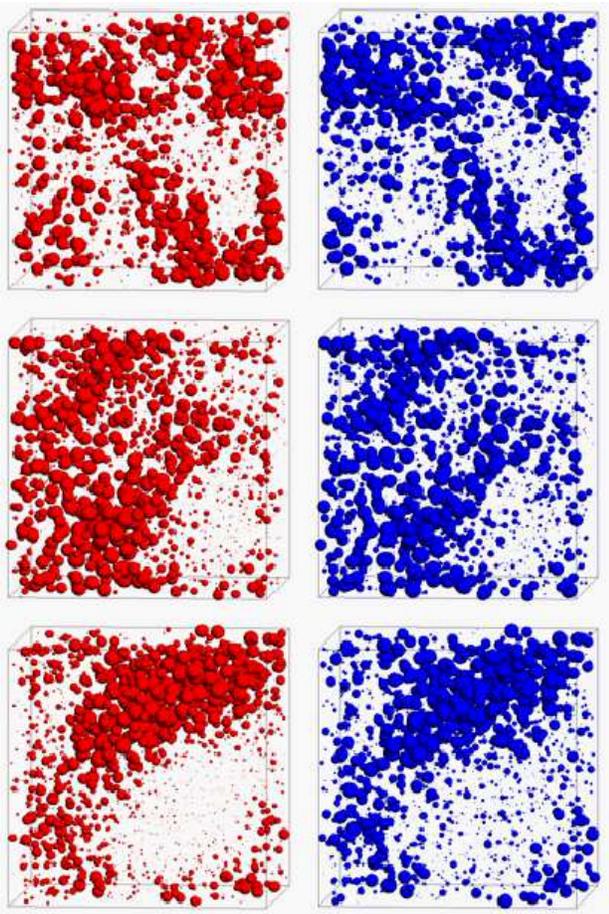}}
\caption{Dynamical heterogeneity (left panels) and structural
  heterogeneity (right panels) in the initial configuration at
  $T=350$~K (top panels), 290~K (middle panels) and 270~K (bottom
  panels).  To make each panel, the values of $\langle r_i^2
  \rangle_{\rm ic}$ (or $\langle u_i \rangle_{\rm ic}$), evaluated at
  the time of the maximum of $S_u$, are assigned to each molecule in
  the initial configuration.  These values are sorted and assigned an
  integer rank $R_i$ from 1 to $N$, from smallest to largest.  Each O
  atom is then plotted as a sphere of radius $R_{\rm
    min}\exp\{[(R_i-N)/(1-N)]\log(R_{\rm max}/R_{\rm min})\}$, where
  $R_{\rm max}=0.14$~nm and $R_{\rm min}=0.004$~nm.  The result
  represents the rank of $\langle r_i^2 \rangle_{\rm ic}$ or $\langle
  u_i \rangle_{\rm ic}$ on an exponential scale, such that the largest
  spheres on the left represent the least mobile molecules, and the
  largest on the right the most tightly bound. Note that hydrogen
  atoms are not shown.}
\label{pix}
\end{figure}

\begin{figure}
\centerline {\includegraphics[width=3.4in]{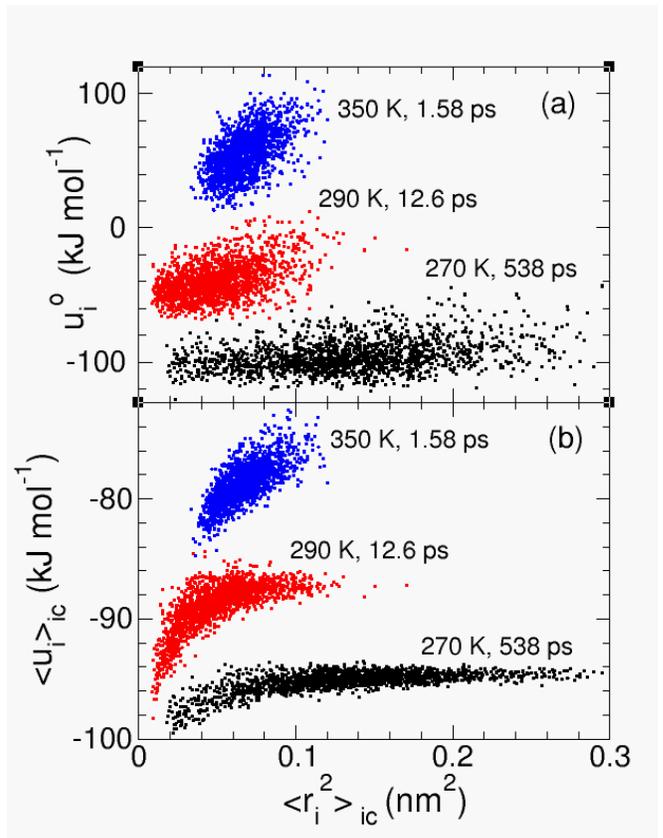}}
\caption{(a) $u_i^{\circ}$ and (b) $\langle u_i \rangle_{\rm ic}$,
  both versus $\langle r_i^2 \rangle_{\rm ic}$ at each $T$.  The times
  indicated, at which $\langle u_i \rangle_{\rm ic}$ and $\langle
  r_i^2 \rangle_{\rm ic}$ are evaluated, correspond to the maxima of
  $S_u$.  To ease comparison, in (a) the cloud for $290$~K has been
  shifted upward by $50$~kJ/mol, and that for $350$~K by 130~kJ/mol;
  in (b) the cloud for $350$~K has been shifted down by 4~kJ/mol.}
\label{scatter}
\end{figure}

The dynamic propensity of each molecule is the value of $\langle r_i^2
\rangle_{\rm ic} = M^{-1}\sum_{k=1}^M r^2(i,k,t)$ when $t$ is on the
order of the structural relaxation time~\cite{CHF04}.  It measures the
propensity for molecule $i$ to undergo a given displacement, given its
starting point in the initial configuration, rather than indicating
what the molecule will do in any particular run.  Here, we extend the
use of the IC ensemble to study structural properties as well as
dynamics.  We focus on $u(i,k,t)$, the contribution of molecule $i$ to
the instantaneous potential energy of the system at time $t$ in run
$k$ of an IC ensemble. Specifically, $u_i=\sum_{j=1}^N\phi_{ij}$,
where $\phi_{ij}$ is the pair potential energy of molecules $i$ and
$j$.  Analogous to $\langle r_i^2\rangle_{\rm ic}$ we define $\langle
u_i \rangle_{\rm ic} = M^{-1}\sum_{k=1}^M u(i,k,t)$.  At a fixed $t$,
$\langle u_i \rangle_{\rm ic}$ measures the propensity of molecule $i$
to have a given value of the potential energy, given its starting
point in the initial configuration.

First we test for the occurrence of DH, by evaluating $\langle
r_i^2\rangle_{\rm ic}$ for each molecule as a function of $t$, and
examining the spatial arrangement of this quantity via a cluster
analysis~\cite{VG04,Starr06}.  For reasons that will become clear
below, we focus on the least mobile molecules, specifically the subset
having the lowest $10\%$ of $\langle r_i^2\rangle_{\rm ic}$ values.
Clusters are defined by the rule that two molecules of this subset
that are also within $0.35$~nm of one another (the position of the
first minimum of the O-O radial distribution function) in the initial
configuration are assigned to the same cluster.  The number-averaged
mean cluster size $S_r$ is then found from $(1/N_c)\sum_nnC(n)$, where
$C(n)$ is the number of clusters of size $n$, and $N_c$ is the total
number of clusters.  Fig.~\ref{ngp}(c) shows the $t$ dependence of
$S_r$ at each $T$.  At small $t$, $S_r$ has a value consistent with a
random choice of $10\%$ of the molecules, approximately $S_r=1.27$,
indicating no spatial correlations in $\langle r_i^2\rangle_{\rm ic}$.
However, at intermediate times corresponding to the onset of
structural relaxation, a maximum occurs, indicating significant
clustering of the least mobile molecules.  At large $t$, $S_r$ begins
its return to the uncorrelated value, and the DH dissolves.  The
morphology of the DH observed near the maxima in Fig.~\ref{ngp}(c) is
illustrated in the left panels of Fig.~\ref{pix}.  The correlated
domains of larger spheres in Fig.~\ref{pix} indicate the locations of
the large clusters that generate the maxima in Fig.~\ref{ngp}(c).

We then carry out exactly the same analysis, but using the lowest
$10\%$ of $\langle u_i \rangle_{\rm ic}$ values; this selects the
subset of molecules with the greatest propensity to be ``tightly
bound.''  The mean cluster size for this subset, $S_u$, shows a
similar behavior to $S_r$ [Fig.~\ref{ngp}(d)].  Note that in the limit
$t\to 0$, we have $\langle u_i \rangle_{\rm ic} \to u_i^{\circ}$,
where $u_i^{\circ}$ is the instantaneous potential energy of each
molecule in the initial configuration.  The behavior of $S_u$ at small
$t$ confirms that $u_i^{\circ}$ shows essentially no spatial
correlation at any $T$.  And yet, on the same time scale as the
signature of DH is observed in $S_r$, a maximum also occurs in $S_u$.
The spatial correlations of molecular potential energies (i.e. {\it
  structural} heterogeneities) that generate the maxima in $S_u$ are
illustrated in Fig.~\ref{pix}.  The morphology of the two types of
heterogeneity shown in Fig.~\ref{pix} is strikingly similar: there is
a clear correlation between regions with a propensity to be the least
mobile, and a propensity to be tightly bound.  It is worth bearing in
mind that these two types of heterogeneity are defined independently.
Neither the time scale on which the structural heterogeneity occurs,
nor the definition of the structural clusters depends on dynamical
information.

The absence of a useful correlation between $u_i^{\circ}$ and the
dynamic propensity is demonstrated by a scatter plot of $u_i^{\circ}$
against $\langle r_i^2\rangle_{\rm ic}$, evaluated at the time of the
maximum of $S_u$ at each $T$ [Fig.~\ref{scatter}(a)].  The best
correlation is found at high $T$, but even here some molecules in the
lowest $10\%$ of $u_i^{\circ}$ have values of $\langle
r_i^2\rangle_{\rm ic}$ comparable or even larger than the mean of
$\langle r_i^2\rangle_{\rm ic}$.  At lower $T$ the correlation only
gets worse: molecules with the lowest values of $u_i^{\circ}$ are
found across the entire spectrum of $\langle r_i^2\rangle_{\rm ic}$
values.  We have also tested if the correlation improves if we replace
$u_i^{\circ}$ in Fig.~\ref{scatter}(a) with its value in the inherent
structure of the initial configuration, but it does not.

Fig.~\ref{scatter}(b) shows a scatter plot of $\langle u_i
\rangle_{\rm ic}$ versus $\langle r_i^2\rangle_{\rm ic}$ at the time
of the maximum of $S_u$ at each $T$.  Here we see the reason why we
have focussed on the least mobile and most tightly bound molecules: it
is at the lower end of these scatter plots that the points are most
easily distinguished from the overall population.  In contrast, the
correlation between the most mobile and least tightly bound molecules
is little improved over that in Fig.~\ref{scatter}(a).  We have also
examined the clusters formed by the $10\%$ most mobile and least
tightly bound subsets of molecules.  The most mobile molecules also
form clusters at intermediate times, although the strength of the
effect is weaker than for the least mobile molecules.  Interestingly,
the least tightly bound subset shows a {\it decreasing} tendency to
cluster as $T$ decreases, with the maximum in $S_u$ becoming difficult
to discern at the lowest $T$.  Whether this is a generic behavior, or
a feature of this particular network-forming liquid requires further
research, which is currently underway.  One possibility is that
molecular hopping is emerging as an important mode of transport at low
$T$, which might weaken the energy-dynamics correlation for the most
mobile molecules~\cite{PG99}.

We can therefore make the following statements about the relationship
of local potential energy and local dynamics in this system: Despite
the absence of a correlation between the instantaneous potential
energy of a molecule and its subsequent displacement, a molecule that
has a propensity to be tightly bound (as evaluated within the IC
ensemble) also has a propensity to be among the least mobile. On the
other hand, a propensity to be mobile and a propensity to be loosely
bound do not correlate at low $T$, indicating that other factors are
controlling dynamics at this end of the spectrum.  The IC ensemble
thus makes possible a dissection of the energy-dynamics relationship,
under conditions where no conclusions can be drawn from instantaneous
structural information.  The approach also exposes the morphology of
the structural heterogeneity present in a single instantaneous
configuration.  As Ref.~\cite{CHF04} first showed, averaging over the
IC ensemble therefore acts as a powerful amplifier of the subtle
configurational signal that must exist instantaneously, but which is
overwhelmed by the ``noise'' imposed by molecular momenta.  While we
have only addressed the behavior of the potential energy in this work,
we expect that valuable information concerning many supercooled
liquids can be extracted from a wide range of configurational
quantities (e.g local volume, local symmetry) when evaluated at the
local level in the IC ensemble.

Our work also demonstrates that the emergence and growth of
heterogeneity in a glass forming liquid is not solely a dynamical
phenomena, but also structural, and that the two types of heterogeneity
can be analyzed within the same framework by using the IC ensemble.
In particular, we note that the size of the structural and dynamical
heterogeneities observed here are approaching the system size at our
lowest $T$. It may be that the ability of the IC ensemble to discern
subtle spatial correlations will help clarify the nature of the
growing length scales of dynamic and structural heterogeneity as $T\to
T_g$.

We thank R. Bowles, P. Kusalik, and F. Starr for useful input;
F. Sciortino for the code to find the inherent structures of the
initial configurations; and especially P. Harrowell for stimulating
our interest in this topic. Financial and computing support was
provided by ACEnet, NSERC, AIF, CFI and the CRC program.


\begin{thebibliography}{99}

\bibitem{E00} M.D. Ediger, Annu. Rev. Phys. Chem. {\bf 51}, 99 (2000).

\bibitem{G00} S.C. Glotzer, Journal of Non-Crystalline Solids {\bf
    274}, 342 (2000).

\bibitem{CHF04} A. Widmer-Cooper, P. Harrowell and H. Fynewever,
  Phys. Rev. Lett. {\bf 93}, 135701 (2004).

\bibitem{CH05} A. Widmer-Cooper and P. Harrowell,
  J. Phys. Condens. Matter {\bf 17}, S4025 (2005).

\bibitem{DH03} B. Doliwa and A. Heuer, Phys. Rev. Lett. {\bf 91},
  235501 (2003); Phys. Rev. E {\bf 67}, 031506 (2003).

\bibitem{SNR81} P.J. Steinhardt, D.R. Nelson and M. Ronchetti,
  Phys. Rev. Lett. {\bf 47}, 1297 (1981).

\bibitem{H82} Y. Hiwatari, J. Chem. Phys. {\bf 76}, 5502 (1982).

\bibitem{JA88} H. Jonsson and H.C. Andersen, Phys. Rev. Lett. {\bf
    60}, 2295 (1988).

\bibitem{Starr02} F.W. Starr, S. Sastry, J.F. Douglas and
  S.C. Glotzer, Phys. Rev. Lett. {\bf 89}, 125501 (2002).

\bibitem{LT06} I. Ladadwa and H. Teichler, Phys. Rev. E {\bf 73},
  031501 (2006).

\bibitem{JP05} T.S. Jain and J.J. de Pablo, J. Chem. Phys. {\bf 122},
  174515 (2005).

\bibitem{CSW05} J.C. Conrad, F.W. Starr and D.A. Weitz,
  J. Phys. Chem. B {\bf 109}, 21235 (2005).

\bibitem{CHcm} A. Widmer-Cooper and P. Harrowell, cond-mat/0511690.

\bibitem{DSZ02} M. Dzugutov, S.I. Simdyankin and F.H.M. Zetterling,
  Phys. Rev. Lett. {\bf 89}, 195701 (2002).

\bibitem{ST06} H. Shintani and H. Tanaka, Nature Physics {\bf 2}, 200
  (2006).

\bibitem{CH06} A. Widmer-Cooper and P. Harrowell,
  Phys. Rev. Lett. {\bf 96}, 185701 (2006).

\bibitem{D99} C. Donati, S.C. Glotzer, P.H. Poole, W. Kob and
  S.J. Plimpton, Phys. Rev. E {\bf 60}, 3107 (1999).

\bibitem{NS04} E. La Nave and F. Sciortino, J. Phys. Chem. B {\bf
    108}, 19663 (2004).

\bibitem{Wales} D. Wales, ``Energy Landscapes'', Cambridge University
  Press, New York (2003).

\bibitem{st2} F.H. Stillinger and A. Rahman, J. Chem. Phys. {\bf 60},
  1545 (1974).

\bibitem{SPES97} F. Sciortino, P.H. Poole, U. Essmann and
  H.E. Stanley, Phys. Rev. E {\bf 55}, 727 (1997).

\bibitem{PG99} D. Paschek and A. Geiger, J. Phys. Chem. B {\bf 103},
  4139 (1999).

\bibitem {PSS05} P.H. Poole, I. Saika-Voivod and F. Sciortino,
  J. Phys. Condens. Matter {\bf 17}, L431 (2005).

\bibitem{BPScm} S.R. Becker, P.H. Poole and F.W. Starr,
  Phys. Rev. Lett., in press (2006).

%\bibitem{beren} H.J.C. Berendsen, J.P.M. Postma, W.F. van~Gunsteren,
%  A. DiNola and J.R. Haak.  J. Chem. Phys. {\bf 81}, 3684 (1984).
\bibitem{beren} H.J.C. Berendsen, et al., J. Chem. Phys. {\bf 81},
  3684 (1984).

\bibitem{VG04} M. Vogel and S.C. Glotzer, Phys. Rev. Lett. {\bf 92},
  255901 (2004); Phys. Rev. E {\bf 70}, 061504 (2004).

\bibitem{Starr06} M.G. Mazza, N. Giovambattista, F.W. Starr and
  H.E. Stanley, Phys. Rev. Lett. {\bf 96}, 057803 (2006).

\end{thebibliography}
\end{document}